\documentclass[12pt]{article}
\begin{document}
\title{The effect of annealing on the nonlinear viscoelastic response 
of isotactic polypropylene}

\author{Aleksey D. Drozdov and Jesper deClaville Christiansen\\
Department of Production\\
Aalborg University\\
Fibigerstraede 16\\
DK--9220 Aalborg, Denmark}
\date{}
\maketitle

\begin{abstract}
Three series of tensile relaxation tests are performed on isotactic 
polypropylene at room temperature in the vicinity of the yield point.
In the first series of experiments, injection-molded samples are
used without thermal pre-treatment.
In the second and third series, prior to testing
the specimens are annealed  at 130 $^{\circ}$C 
for 4 and 24 hours, respectively.

Constitutive equations are derived for the time-dependent
response of semicrystalline polymers at isothermal loading
with small strains.
A polymer is treated as an equivalent temporary network 
of macromolecules bridged by junctions (physical cross-links, 
entanglements and crystalline lamellae).
Under loading, junctions slip with respect to their positions
in the bulk material (which reflects the viscoplastic behavior), 
whereas chains separate from their junctions and merge with 
new ones at random times (which reflects the viscoelastic response).
The network is thought of as an ensemble of meso-regions (MR) 
with various activation energies for detachment of chains 
from temporary nodes.

Adjustable parameters in the stress--strain relations are found
by fitting observations.
Experimental data demonstrate that the shape of the relaxation spectrum 
(characterized by the distribution of MRs with various potential energies)
is independent of mechanical factors, but is altered at annealing.
For specimens not subjected to thermal treatment, 
the growth of longitudinal strain does not affect the volume fraction 
of active MRs and the attempt rate for detachment of chains 
from their junctons.
For annealed samples, the concentration of active MRs
increases and the attempt rate decreases with strain.
These changes in the time-dependent response 
are attributed to broadening of the distribution of strengths of
crystalline lamellae at annealing.
\end{abstract}

\section{Introduction}

This paper is concerned with the influence of annealing at
an elevated temperate on the nonlinear viscoelastic response
of isotactic polypropylene (iPP) at room temperature.
The objective of this study is three-fold:
\begin{enumerate}
\item
to report experimental data in tensile relaxation tests 
on specimens annealed for various amounts of time
at strains in the vicinity of the yield point,

\item
to derive stress--strain relations for the nonlinear
viscoelastic behavior of a semicrystalline polymer
at isothermal uniaxial deformation,

\item
to assess the effect of annealing on the time-dependent
response of iPP in terms of the constitutive model.
\end{enumerate}
Isotactic polypropylene is chosen for the analysis because of
numerous applications of this polymer in industry 
(oriented films for packaging,
reinforcing fibres,
nonwoven fabrics,
blends with thermoplastic elastomers, etc.).
The goal of this study is to establish some correlations between
mechanical properties, morphology and processing conditions
(annealing at an elevated temperature) for injection-molded specimens.
For a review of previous works on this subject, 
see \cite{KB97} and the bibliography therein.

The nonlinear viscoelastic response of polypropylene was analyzed 
by Ward and Wolfe \cite{WW66}, see also \cite{WH93},
and Smart and Williams \cite{SW72} three decades ago,
and, more recently, by Ariyama \cite{Ari93a,Ari93b,Ari96,AMK97},
Wortmann and Schulz \cite{WS94,WS95},
Ibhadon \cite{Ibh96},
Tomlins \cite{Tom96},
Read and Tomlins \cite{RT97a,RT97b},
Dutta and Edward \cite{DE97}
and Tomlins and Read \cite{TR98}.

The effect of physical aging (annealing at an elevated temperature
followed by quench to ambient temperature) on the time-dependent
behavior of PP was studied by Struik \cite{Str78,Str87},
Chai and McCrum \cite{CM80},
Ibhadon \cite{Ibh96},
Tomlins \cite{Tom96},
Read and Tomlins \cite{RT97a,RT97b}
and Tomlins and Read \cite{TR98}.

Dynamic mechanical analysis shows that the loss tangent 
of iPP demonstrates two pronounced maxima 
being plotted versus temperature \cite{And99,SSE99}.
The first maximum ($\beta$--transition in the interval 
between $T=-20$ and $T=10$ $^{\circ}$C) is associated with 
the glass transition in the most mobile part of the amorphous phase, 
whereas the other maximum ($\alpha$--transition in the interval 
between $T=70$ and $T=110$ $^{\circ}$C) is attributed to the
glass transition in the remaining part of the amorphous phase.
This conclusion is confirmed by DSC (differential scanning
calorimetry) traces for quenched PP that reveal an endoterm at $T=70$ 
$^{\circ}$C which can be ascribed to thermal activation of  
amorphous regions with restricted mobility under heating \cite{SSE99}.

Isotactic polypropylene exhibits three different crystallographic forms:
monoclinic $\alpha$ crystallites,
(pseudo) hexagonal $\beta$ structures,
orthorhombic $\gamma$ polymorphs,
and ``smectic" mesophase (arrays of chains with a better order
in the longitudinal than in transverse chain direction).
For a detailed review of iPP polymorphs, the reader is referred to 
the survey \cite{Bic98}.
At rapid cooling of the melt (which is typical of injection molding),
$\alpha$ crystallites and smectic mesophase are mainly developed,
whereas metastable $\beta$ and $\gamma$ structures arise as
minority components.
Crystallization of $\beta$ forms occurs either under stresses
or with the help of $\beta$-nucleating agents added to the melt 
\cite{LGS01,LVS02}.
Formation of $\gamma$ polymorph requires high pressure
for commercial grades of iPP, while it can observed 
at atmospheric pressure in isotactic polypropylene 
with low molecular weight \cite{AKG99}.
A unique feature of $\alpha$ structures in iPP is the 
lamellar crosshatching: development of transverse lamellae 
in spherulites that are oriented in the direction perpendicular to 
the direction of radial lamellae \cite{IS00}.

Scanning electron microscopy \cite{AGU95,LFG99},
atomic force microscopy \cite{CCG98}
and X-ray diffraction \cite{SSE99,LGS01,LVS02}
reveal that in injection-molded specimens
$\alpha$ spherulites have the characteristic size of the order
of 100 $\mu$m and they contain crystalline lamellae 
with thickness of 10 to 20 nm.
The amorphous phase is located  between spherulites
and inside the spherulites between lamellae.
It consists of (i) relatively mobile chains between spherulites
and between radial lamellae inside spherulites,
and (ii) severely restricted chains in the regions bounded 
by radial and tangential lamellae in $\alpha$ spherulites.

Annealing of injection-molded iPP at an elevated temperature
results in 
(i) secondary crystallization of a part of the amorphous phase, 
(ii) thickening of radial lamellae,
(iii) development of subsidiary lamellae,
(iv) formation of lamellar superstructure,
and (v) growth of the crystal perfection  \cite{MHY00}.
Other changes in the crystalline morphology of iPP
driven by thermal treatment are the subject of debate.
Some researchers \cite{LGS01,LVS02,LFG99,CCG98}
conclude that the fraction of $\beta$ spherulites 
increases at annealing in the interval of 
temperatures between 110 and 140 $^{\circ}$C,
which enhances ductility of iPP and improves its 
impact properties.
According to other authors \cite{SSE99,IS00,AW01}, 
annealing of iPP induces transformation of the smectic phase 
into monoclinic $\alpha$ spherulites without noticeable 
development of $\beta$ polymorph.

Mechanical loading results in inter-lamellar separation,
rotation and twist of lamellae,
fine and coarse slip of lamellar blocks and
their fragmentation \cite{SSE99}.
Straining of iPP specimens causes chain slip through 
the crystals, sliding and breakage of tie chains
and activation of restricted amorphous regions driven
by lamellar disintegration.
In the post-yield region, these changes in the micro-structure
imply cavitation, breakage of crystalls, 
and formation of fibrills \cite{ZBC99}.

It is hard to believe that these morphological transformations 
in iPP can be adequately described by a constitutive model 
with a small number of adjustable parameters.
To develop stress--strain relations,
we apply a method of ``homogenization of micro-structure"
\cite{BKR02}.
According to this approach, an equivalent phase 
is introduced whose deformation captures 
essential features of the response of a semicrystalline 
polymer with a complicated micro-structure.
In this study, an amorphous phase is chosen as 
the equivalent phase because of the following reasons:
\begin{enumerate}
\item
The viscoelastic response of semicrystalline polymers 
is conventionally associated with rearrangement of chains 
in amorphous regions \cite{CCG98}.

\item
Sliding of tie chains along and their detachment from lamellae 
play the key role in the time-dependent response of
semicrystalline polymers  \cite{NT99,NT00}.

\item
The viscoplastic flow in semicrystalline polymers
is assumed to be ``initiated in the amorphous phase 
before transitioning into the crystalline phase" \cite{MP01}.

\item
Conventional models for  polyethylene \cite{BKR02},
polypropylene  \cite{SW96}
and poly(ethylene terephthalate) \cite{BSL00}
treat these polymers as networks of macromolecules.
\end{enumerate}

Above the glass transition temperature for the mobile 
amorphous phase, isotactic polypropylene is thought of 
as a network of chains bridged by junctions.
Deformation of a specimen induces slip of junctions 
with respect to their positions in the bulk material.
Sliding of junctions reflects slippage of tie molecules
along lamellae and fine slip of lamellar blocks
which are associated with the viscoplastic behavior
of a semicrystalline polymer.

With reference to the concept of transient networks 
\cite{GT46,Yam56,Lod68,TE92},
the viscoelastic response of iPP is modelled as 
separation of active chains from their junctions 
and attachment of dangling chains to temporary nodes.
The network of macromolecules is assumed to be strongly 
inhomogeneous (this heterogeneity reflects the effect of 
spherulites on rearrangement of surrounding chains),
and it is treated as an ensemble of meso-regions (MR) with various 
potential energies for detachment of active strands.
Two types of MRs are distinguished: (i) active domains where
strands separate from junctions as they are thermally agitated
(these MRs model a mobile part of the amorphous phase), 
and (ii) passive domains where detachment of
chains from junctions is prevented.
Passive MRs are associated with a part of the amorphous phase 
whose mobility is restricted by (i) radial and tangential lamellae 
and (ii) surrounding macromolecules (because of density fluctuations
in the amorphous phase).

Separation of active chains from temporary nodes 
is treated as a thermally-activated process
whose rate obeys the Eyring equation \cite{KE75} with a
strain-dependent attempt rate.
An increase in the relaxation rate of amorphous polymers 
at straining is conventionally attributed to the 
mechanically-induced growth of ``free volume" 
between macromolecules which, in turn, 
implies an increase in their mobility \cite{KE87,LK92,OK95}.

Stretching of a specimen results in 
(i) a mechanically-induced changes in the rate of detachmens
of strands in active MRs and
(ii) an increase in the concentration of active MRs.
The latter is ascribed to (i) partial release of the amorphous phase
in passive meso-domains driven by fragmentation of lamellae
and (ii) breakage of van der Waals links between compactly 
packed chains in meso-domains with higher density.

The exposition is organized as follows.
Section 2 is concerned with the description of the experimental
procedure.
Kinetic equations for sliding of junctions and 
reformation of active strands are developed in Section 3.
Constitutive equations for uniaxial deformation
are derived in Section 4.
In Section 5 these relations are applied to fit experimental data.
A brief discussion of our findings is presented in Section 6.
Some concluding remarks are formulated in Section 7.

\section{Experimental procedure}

Isotactic polypropylene (Novolen 1100L) was supplied by BASF (Targor).
ASTM dumbell specimens were injection molded 
with length 14.8 mm, width 10 mm and height 3.8 mm.
Three series of tests were performed.
In the first series, the samples were used as received
without thermal pre-treatment.
In the second series, the specimens were annealed in an oven
at the temperature 130 $^{\circ}$C for 4 h and slowly cooled by air.
In the third series of experiments, the specimens were annealed 
at the same temperature for 24 h and cooled by air.

Differential scanning calorimetry measurements were carried out
on STA 449/Netzsch apparatus at the heating rate 5 K/min.
The specimens with weight of about 15 mg
were tested in Al$_{2}$O$_{2}$ pans covered by lid.
The thermal analyzer was calibrated with 7 references ranging
from In to Ni.
The specific  enthalpy of melting, $\Delta H_{\rm m}$, 
equals 86.9,  98.1 and 101.7 J/g, for non-annealed specimens,
specimens annealed for 4 h, and specimens annealed for 24 h,
respectively.
With reference to \cite{Wun80}, we accept the value 209 J/g
as the enthalpy of fusion for a fully crystalline polypropylene.
The degree of crystallinity, $\kappa_{\rm c}$, is estimated as 
41.6\%, 46.8 \% and 48.7 \% for specimens not subjected to thermal
treatment, samples annealed for 4 h, and specimens 
annealed for 24 h, respectively.

Although the degree of crystallinity changes rather weakly
(but consistently) with an increase in the annealing time,
the shape of DSC curves is noticeably altered in the interval
of temperatures between 120 and 160 $^{\circ}$C.
DSC traces depicted in Figure 1 are similar to those found
by other authors, see, e.g.,  \cite{LGS01,IS00}.
Labour et al. \cite{LGS01} attributed the growth of the
low-temperature shoulder on the melting curve to the 
$\alpha\to\beta$ transition at annealing.
According to Iijima and Strobl \cite{IS00}, the endothermal
contribution on the low-temperature side of the DSC trace
indicates the presence of $\alpha$ crystallites with varying
stability.

Uniaxial tensile relaxation tests were performed 
at room temperature on a testing machine 
Instron--5568 equipped with electro-mechanical sensors 
for the control of longitudinal strains in the active zone of samples 
(the distance between clips was about 50 mm).
The tensile force was measured by the standard loading cell.
The engineering stress $\sigma$ was determined
as the ratio of the axial force to the cross-sectional area
of the specimens in the stress-free state.

Any series of mechanical experiments included 9 relaxation tests 
at the longitudinal strains
$\epsilon_{1}=0.02$,
$\epsilon_{2}=0.04$,
$\epsilon_{3}=0.06$,
$\epsilon_{4}=0.08$,
$\epsilon_{5}=0.10$,
$\epsilon_{6}=0.12$,
$\epsilon_{7}=0.14$,
$\epsilon_{8}=0.16$,
$\epsilon_{9}=0.18$,
which corresponded to the domain of nonlinear viscoelasticity,
sub-yield and post-yield regions for isotactic polypropylene 
(the yield strain, $\epsilon_{\rm y}$, was estimated by the supplier as 0.13).
Mechanical tests were carried out at least one day after annealing
of specimens to avoid the influence of physical aging on the
time-dependent response of iPP.

Each relaxation test was performed on a new sample.
No necking of specimens was observed in experiments
(except for the test with $\epsilon_{9}=0.18$ on a specimen
not subjected to thermal treatment,
which was excluded from the consideration).
In the $k$th relaxation test ($k=1,\ldots,9$), a specimen was loaded 
with the cross-head speed 5 mm/min (that roughly corresponded 
to the strain rate  $\dot{\epsilon}_{0}=0.05$ min$^{-1}$) 
up to the longitudinal strain $\epsilon_{k}$, which was preserved
constant during the relaxation time $t_{\rm r}=20$ min.

The engineering stresses, $\sigma$, at the beginning of the relaxation
tests are plotted in Figure 2 together with the stress--strain curves for
the specimens strained up to $\epsilon_{9}=0.18$.
The figure demonstrates fair repeatability of experimental data.

Figure 2 shows that annealing for 4 h results in a pronounced increase
in stress compared to virgin specimens.
The growth of the annealing time implies a decrease in stress 
in the sub-yield region.
The discrepancy between the stress--strain curves for specimens 
annealed for 4 and 24 h practically disappears in the post-yield domain.

Despite the coincidence of the stress--strain diagrams 
in the post-yield region for specimens annealed for 4 and 24 h, 
these samples demonstrate a noticeably different
necking behavior. 
Necking of specimens not subjected to thermal pre-treatment
occurs at the strain $\epsilon_{\rm n}=0.18$,
necking of specimens annealed for 4 h takes place at the
strain $\epsilon_{\rm n}=0.25$,
whereas no necking is observed for specimens annealed for 24 h
at stretching up to the strain $\epsilon=0.30$.

The longitudinal stress, $\sigma$, is plotted versus  the logarithm 
($\log=\log_{10}$) of time $t$ (the initial instant $t=0$ corresponds 
to the beginning of the relaxation process) in Figures 3 to 11.
These figures demonstrate that the time of annealing strongly
affects the shape of relaxation curves (especially, in the sub-yield 
region, see Figures 3 and 4).
For any strain $\epsilon>0.02$, stresses in the annealed specimens 
exceed those in the samples not subjected to thermal pre-treatment.
In the sub-yield domain ($\epsilon<0.1$) stresses in the specimens
annealed for 4 h are higher than stresses in the samples annealed
for 24 h.

Our aim now is to develop constitutive equations 
for the time-dependent behavior of a semicrystalline
polymer to be employed for the quantitative analysis of
these observations.

\section{A micro-mechanical model}
 
A semicrystalline polymer is treated as a temporary network of 
chains bridged by junctions.
The network is modelled as an ensemble of meso-regions with
various strengths of interaction between macromolecules.
Two types of meso-domains are distinguished: passive and active.
In passive MRs, inter-chain interaction prevents detachment of 
chains from junctions, which implies that all  nodes in these domains
are permanent.
In active MRs, active strands (whose ends are connected to 
contiguous junctions) separate from the temporary junctions 
at random times when they are thermally agitated.
An active chain whose end detaches from a junction is transformed
into a dangling chain.
A dangling chain  returns into the active state
when its free end captures a nearby junction at a random instant.

Denote by $X$ the average number of active strands per unit mass
of a polymer, by $X_{\rm a}$ the number of strands merged 
with the network in active MRs, and by $X_{\rm p}$ the number 
of strands connected to the network in passive MRs.
Under stretching some crystalline lamellae (restricting mobility
of chains in passive MRs) break,
which results in a growth of the number of strands to be rearranged.
As a consequence, the number of strands in active MRs increases
and the number of strands in passive meso-domains decreases.
This implies that the quantities $X_{\rm a}$ and $X_{\rm p}$ 
become functions of the current strain, $\epsilon$, that
obey the conservation law
\begin{equation}
X_{\rm a}(\epsilon) +X_{\rm p}(\epsilon) =X.
\end{equation}
Rearrangement of strands in active MRs is thought of as
a thermally activated process.
The rate of detachment of active strands from their junctions
in a MR with potential energy $\bar{\omega}$ in the stress-free 
state of a specimen is given by the Eyring equation \cite{KE75}
\[
\Gamma=\Gamma_{\rm a}\exp\biggl (-\frac{\bar{\omega}}{k_{\rm B}T}\biggr ),
\]
where $k_{\rm B}$ is Boltzmann's constant, 
$T$ is the absolute temperature, 
and the pre-factor $\Gamma_{\rm a}$ is independent
of energy $\bar{\omega}$ and temperature $T$.
Introducing the dimensionless activation energy
$\omega=\bar{\omega}/(k_{\rm B}T_{0})$, 
where $T_{0}$ is a reference temperature,
and disregarding the effects of small increments of
temperature, $\Delta T=T-T_{0}$, on the rate of separation, $\Gamma$,
we arrive at the formula
\begin{equation}
\Gamma=\Gamma_{\rm a}\exp (-\omega).
\end{equation}
We suppose that Eq. (2) remains valid for an arbitrary
loading process, provided that the attempt rate, $\Gamma_{\rm a}$,
is a function of the current strain, $\Gamma_{\rm a}=\Gamma_{\rm a}(\epsilon)$.

The distribution of active MRs with various potential 
energies is described by the probability density $p(\omega)$ that equals the
ratio of the number, $N_{\rm a}(\epsilon,\omega)$, of active meso-domains 
with energy $\omega$ to the total number of active MRs,
\begin{equation}
N_{\rm a}(\epsilon,\omega)=X_{\rm a}(\epsilon)p(\omega).
\end{equation}
The distribution function for potential energies of active MRs, 
$p(\omega)$, is assumed to be strain-independent.

The ensemble of active meso-domains is described by the
function $n_{\rm a}(t,\tau,\omega)$ that equals the number of
active strands at time $t$ (per unit mass) belonging to
active MRs with potential energy $\omega$ that have last been 
rearranged before instant $\tau\in [0,t]$.
In particular, $n_{\rm a}(0,0,\omega)$ is the number (per unit mass)
of active strands in active MRs with potential 
energy $\omega$ in a stress-free medium,
\begin{equation}
n_{\rm a}(0,0,\omega)=N_{\rm a}(0,\omega), 
\end{equation}
and $n_{\rm a}(t,t,\omega)$ is the number (per unit mass)
of active strands in active MRs with potential 
energy $\omega$ in the deformed medium at time $t$
(the initial time $t=0$ corresponds to the instant when
external loads are applied to the polymer),
\begin{equation}
n_{\rm a}(t,t,\omega)=N_{\rm a}(\epsilon(t),\omega). 
\end{equation}
The amount
\[ \frac{\partial n_{\rm a}}{\partial \tau}(t,\tau,\omega)\biggl |_{t=\tau} d\tau \]
equals the number (per unit mass) of dangling strands in 
active MRs with potential energy $\omega$  that merge 
with the network within the interval $[\tau,\tau+d\tau ]$,
and the quantity
\[ \frac{\partial n_{\rm a}}{\partial \tau}(t,\tau,\omega) d\tau \]
is the number of there strands that have not detached from
temporary junctions during the interval $[\tau, t]$.
The number (per unit mass) of strands in active MRs that separate (for
the first time) from the network within the interval $[t,t+dt]$ reads
\[ -\frac{\partial n_{\rm a}}{\partial t}(t,0,\omega) dt, \]
whereas the number (per unit mass) of strands in active MRs 
that merged with the network during the interval $[\tau,\tau+d\tau ]$
and, afterwards, separate from the network
within the interval $[t,t+dt]$ is given by
\[ -\frac{\partial^{2} n_{\rm a}}{\partial t\partial \tau}(t,\tau,\omega) dt d\tau. \]
The rate of detachment, $\Gamma$, equals the ratio of
the number of active strands that separate from the network per unit
time to the current number of active strands.
Applying this definition to active strands that merged with the network
during the interval $[\tau,\tau+d\tau ]$
and separate from temporary junctions within the interval $[t,t+dt]$, 
we find that
\begin{equation}
\frac{\partial^{2} n_{\rm a}}{\partial t\partial \tau}(t,\tau,\omega)=-
\Gamma(\epsilon(t),\omega)
\frac{\partial n_{\rm a}}{\partial \tau}(t,\tau,\omega).
\end{equation}
Changes in the function $n_{\rm a}(t,0,\omega)$ are governed
by two processes at the micro-level: 
(i) detachment of active strands from temporary nodes,
and (ii) mechanically-induced activation of passive MRs.
The kinetic equation for this function reads
\begin{equation}
\frac{\partial n_{\rm a}}{\partial t}(t,0,\omega)=-
\Gamma(\epsilon(t),\omega) n_{\rm a}(t,0,\omega)
+\frac{\partial N_{\rm a}}{\partial \epsilon}(\epsilon(t),\omega)
\frac{d\epsilon}{dt}(t).
\end{equation}
The solution of Eq. (7) with initial condition (4) is given by
\begin{eqnarray}
n_{\rm a}(t,0,\omega) &=& N_{\rm a}(0,\omega)\exp \biggl [ -
\int_{0}^{t} \Gamma(\epsilon(s),\omega)ds \biggr ]
\nonumber\\
&&+\int_{0}^{t} \frac{\partial N_{\rm a}}{\partial \epsilon}
(\epsilon(\tau),\omega) \frac{d\epsilon}{dt}(\tau)
\exp \biggl [ - \int_{\tau}^{t} \Gamma(\epsilon(s),\omega)ds \biggr ]
d\tau .
\end{eqnarray}
It follows from Eq. (6) that 
\begin{equation}
\frac{\partial n_{\rm a}}{\partial \tau}(t,\tau,\omega)
=\varphi(\tau,\omega)
\exp \biggl [ - \int_{\tau}^{t} \Gamma(\epsilon(s),\omega)ds \biggr ],
\end{equation}
where
\begin{equation}
\varphi(\tau,\omega)
=\frac{\partial n_{\rm a}}{\partial \tau}(t,\tau,\omega)\biggr |_{t=\tau}.
\end{equation}
To determine the function $\varphi(t,\omega)$, we use the identity
\begin{equation} 
n_{\rm a}(t,t,\omega)=n_{\rm a}(t,0,\omega)
+\int_{0}^{t} \frac{\partial n_{\rm a}}{\partial \tau}(t,\tau,\omega) d\tau.
\end{equation}
Equations (5) and (11) imply that
\begin{equation}
n_{\rm a}(t,0,\omega)
+\int_{0}^{t} \frac{\partial n_{\rm a}}{\partial \tau}(t,\tau,\omega) d\tau
=N_{\rm a}(\epsilon(t),\omega).
\end{equation}
Differentiating Eq. (12) with respect to time and using Eq. (10), we obtain
\[
\varphi(t,\omega)+\frac{\partial n_{\rm a}}{\partial t}(t,0,\omega)
+\int_{0}^{t} \frac{\partial^{2} n_{\rm a}}{\partial t\partial \tau}(t,\tau,\omega)
d\tau=\frac{\partial N_{\rm a}}{\partial \epsilon}(\epsilon(t),\omega)
\frac{d\epsilon}{dt}(t).
\]
This equality together with Eqs. (6), (7) and (11) results in 
\begin{equation}
\varphi(t,\omega) =  \Gamma(\epsilon(t),\omega)n_{\rm a}(t,t,\omega).
\end{equation}
Substituting expression (13) into Eq. (9) and using Eq. (5), we
arrive at the formula
\begin{equation}
\frac{\partial n_{\rm a}}{\partial \tau}(t,\tau,\omega)
=\Gamma(\epsilon(t),\omega)N_{\rm a}(\epsilon(t),\omega)
\exp \biggl [ - \int_{\tau}^{t} \Gamma(\epsilon(s),\omega)ds \biggr ].
\end{equation}
The kinetics of rearrangement of strands in active MRs
is described by Eqs. (2), (3), (8) and (14). 
These relations are determined by (i) the distribution function
$p(\omega)$ for active MRs with various potential energies
$\omega$, (ii) the function $\Gamma_{\rm a}(\epsilon)$ that
characterizes the effect of strains on the attempt rate,
and (iii) the function 
\begin{equation}
\kappa_{\rm a}(\epsilon)=\frac{X_{\rm a}(\epsilon)}{X},
\end{equation}
that reflects mechanically-induced activation of passive MRs.

Rearrangement of strands in active MRs reflects
the viscoelastic response of a semicrystalline polymer.
The viscoplastic behavior is associated with 
the mechanically-induced slippage of junctions 
with respect to their positions in the bulk material.

Denote by $\epsilon_{\rm u}(t)$ the average strain induced by sliding
of junctions between macromolecules (the subscript index ``u" means
that $\epsilon_{\rm u}(t)$ coincides with the residual strain in a 
specimen which is suddenly unloaded at instant $t$).
The elastic strain (that reflects elongation of active strands  in 
a network) is denoted by $\epsilon_{\rm e}(t)$.
The strains $\epsilon_{\rm e}(t)$ and $\epsilon_{\rm u}(t)$ are 
connected with the macro-strain $\epsilon(t)$ by
the conventional formula
\begin{equation}
\epsilon(t)=\epsilon_{\rm e}(t)+\epsilon_{\rm u}(t).
\end{equation}
We adopt the first order kinetics for slippage of junctions with
respect to the bulk material, which implies that the increment of the
viscoplastic strain, $d\epsilon_{\rm u}$, induced by the growth of 
the macro-strain, $\epsilon$, by an increment, $d\epsilon$, 
is proportional to the absolute value of the stress $\sigma$,
\begin{equation}
\frac{d\epsilon_{\rm u}}{d\epsilon}=B |\sigma |\; {\rm sign} 
\Bigl ( \sigma\frac{d\epsilon}{dt}\Bigr ),
\end{equation}
where the pre-factor $B$ is a non-negative function of stress, 
strain  and the strain rate,
\[ B=B\Bigl (\sigma, \epsilon, \frac{d\epsilon}{dt}\Bigr ). \]
The last multiplier in Eq. (17) determines the direction of 
the viscoplastic flow of junctions.
Equation (17) is presented in the form
\begin{equation}
\frac{d\epsilon_{\rm u}}{dt}(t)=B\Bigl (\sigma(t), \epsilon(t),
\frac{d\epsilon}{dt}\Bigr ) | \sigma(t) |\; 
{\rm sign} \Bigl [ \sigma(t) \frac{d\epsilon}{dt}(t) \Bigr ]
\frac{d\epsilon}{dt}(t),
\qquad
\epsilon_{\rm u}(0)=0,
\end{equation}
which implies that the rate of sliding vanishes when
the macro-strain, $\epsilon$, remains constant.

\section{Constitutive equations}

An active strand is modelled as a linear elastic solid with the mechanical energy
\[
w(t)=\frac{1}{2}\mu e^{2}(t), 
\]
where $\mu$ is the average rigidity per strand
and $e$ is the strain from the stress-free state to the deformed state.

For strands belonging to passive meso-domains, the strain $e$
coincides with $\epsilon_{\rm e}$.
Multiplying the strain energy per strand by the number of strands in
passive MRs, we find the mechanical energy of meso-domains 
where rearrangement of chains is prevented by surrounding lamellae,
\begin{equation}
W_{\rm p}(t)=\frac{1}{2}\mu X_{\rm p}(\epsilon(t))\epsilon_{\rm e}^{2}(t). 
\end{equation}
With reference to the conventional theory of temporary networks \cite{TE92},
we assume that stresses in dangling strands totally relax before
these strands merge with the network.
This implies that the reference (stress-free) state of a strand that
is attached to the network at time $\tau$ coincides with
the deformed state of the network at that instant.
For active strands that have not been rearranged until time $t$,
the strain $e(t)$ coincides with $\epsilon_{\rm e}(t)$, 
whereas for active strands that have last been merged with the network 
at time $\tau\in [0,t]$, the strain $e(t,\tau)$ is given by
\[ 
e(t,\tau)=\epsilon_{\rm e}(t)-\epsilon_{\rm e}(\tau).
\]
Summing the mechanical energies of active strands 
belonging to active MRs with various potential energies, $\omega$,
that were rearranged at various instants, $\tau$, we find the
mechanical energy of active meso-domains,
\begin{equation}
W_{\rm a}(t) = \frac{1}{2}\mu \int_{0}^{\infty} d\omega 
 \biggl \{ n_{\rm a}(t,0,\omega)\epsilon_{\rm e}^{2}(t)
+\int_{0}^{t} \frac{\partial n_{\rm a}}{\partial \tau}(t,\tau,\omega)
\Bigl [ \epsilon_{\rm e}(t)-\epsilon_{\rm e}(\tau)\Bigr ]^{2} d\tau \biggr \}.
\end{equation}
The mechanical energy per unit mass of a polymer reads
$W(t)=W_{\rm a}(t)+W_{\rm p}(t)$.
Substituting expressions (19) and (20) into this equality and using Eq. (16),
we arrive at the formula
\begin{eqnarray}
W(t) &=&  \frac{1}{2}\mu \biggl \{  X_{\rm p}(\epsilon(t))
\Bigl (\epsilon(t)-\epsilon_{\rm u}(t) \Bigr )^{2}(t)
+  \int_{0}^{\infty} d\omega  \biggl [ n_{\rm a}(t,0,\omega)
\Bigl (\epsilon(t)-\epsilon_{\rm u}(t) \Bigr )^{2}
\nonumber\\
&& +\int_{0}^{t} \frac{\partial n_{\rm a}}{\partial \tau}(t,\tau,\omega)
\Bigl ( \Bigl ( \epsilon(t)-\epsilon_{\rm u}(t)\Bigr )
-\Bigl ( \epsilon(\tau)-\epsilon_{\rm u}(\tau)\Bigr ) \Bigr )^{2} d\tau \biggr ]
\biggr \}.
\end{eqnarray}
Differentiation of Eq. (21) with respect to time results in
\begin{equation}
\frac{dW}{dt}(t)=\mu \Bigl [ A(t)  \frac{d\epsilon}{dt}(t)
-\frac{1}{2}\Bigl (A_{1}(t)+A_{2}(t)\Bigr )\Bigr ],
\end{equation}
where
\begin{eqnarray}
A(t) &=& X_{\rm p}(\epsilon(t))\Bigl [ \epsilon(t)-\epsilon_{\rm u}(t)\Bigr ]
+\int_{0}^{\infty} d\omega \biggl \{ n_{\rm a}(t,0,\omega)
\Bigl [ \epsilon(t)-\epsilon_{\rm u}(t)\Bigr ]
\nonumber\\
&& +\int_{0}^{t} \frac{\partial n_{\rm a}}{\partial \tau}(t,\tau,\omega)
\Bigl [ \Bigl (\epsilon(t)-\epsilon_{\rm u}(t)\Bigr )
-\Bigl (\epsilon(\tau)-\epsilon_{\rm u}(\tau)\Bigr )\Bigr ]d\tau \biggr \},
\nonumber\\
A_{1}(t) &=& -\frac{\partial X_{\rm p}}{\partial \epsilon}(\epsilon(t))
\frac{d\epsilon}{dt}(t) \Bigl [ \epsilon(t)-\epsilon_{\rm u}(t)\Bigr ]^{2}
-\int_{0}^{\infty} d\omega \biggl \{ \frac{\partial n_{\rm a}}{\partial t}
(t,0,\omega)\Bigl [ \epsilon(t)-\epsilon_{\rm u}(t)\Bigr ]^{2}
\nonumber\\
&& +\int_{0}^{t} \frac{\partial^{2} n_{\rm a}}{\partial t\partial \tau}(t,\tau,\omega)
\Bigl [ \Bigl (\epsilon(t)-\epsilon_{\rm u}(t)\Bigr )
-\Bigl (\epsilon(\tau)-\epsilon_{\rm u}(\tau)\Bigr )\Bigr ]^{2} d\tau \biggr \},
\nonumber\\
A_{2}(t) &=& 2A(t)\frac{d\epsilon_{\rm u}}{dt}(t).
\end{eqnarray}
Bearing in mind Eqs. (1), (3), (5) and (11), we transform the first equality in Eq. (23) 
as follows:
\begin{equation}
A(t) = X \Bigl [ \epsilon(t)-\epsilon_{\rm u}(t)\Bigr ]
-\int_{0}^{\infty} d\omega 
\int_{0}^{t} \frac{\partial n_{\rm a}}{\partial \tau}(t,\tau,\omega)
\Bigl [\epsilon(\tau)-\epsilon_{\rm u}(\tau) \Bigr ]d\tau .
\end{equation}
Substitution of expressions (1), (3), (6) and (7) into the second equality in Eq. (23)
yields
\begin{eqnarray}
A_{1}(t) &=& \int_{0}^{\infty} \Gamma(\epsilon(t),\omega) d\omega
\biggl \{ n_{\rm a}(t,0,\omega) \Bigl [ \epsilon(t)-\epsilon_{\rm u}(t)\Bigr ]^{2}
\nonumber\\
&& +\int_{0}^{t} \frac{\partial n_{\rm a}}{\partial \tau}(t,\tau,\omega)
\Bigl [ \Bigl (\epsilon(t)-\epsilon_{\rm u}(t)\Bigr )
-\Bigl (\epsilon(\tau)-\epsilon_{\rm u}(\tau)\Bigr )\Bigr ]^{2} d\tau \biggr \}.
\end{eqnarray}
For uniaxial loading with small strains at the reference 
temperature $T_{0}$, the Clausius-Duhem inequality reads
\[
T_{0}Q(t)=-\frac{dW}{dt}(t)+\frac{1}{\rho}\sigma(t)\frac{d\epsilon}{dt}(t)
\geq 0,
\]
where $\rho$ is mass density,
and $Q$ is the rate of entropy production per unit mass.
Substitution of expression (22) into this equation implies that
\begin{equation}
T_{0}Q(t)=\frac{1}{\rho}\Bigl [ \sigma(t)-\rho\mu A(t)\Bigr ] \frac{d\epsilon}{dt}(t)
+\frac{1}{2}\Bigl [ A_{1}(t)+A_{2}(t)\Bigr ] \geq 0.
\end{equation}
Because Eq. (26) is to be fulfilled for an arbitrary program 
of straining, $\epsilon=\epsilon(t)$, the expression in the first square 
brackets vanishes.
This assertion together with Eq. (24) results in the stress--strain relation
\begin{equation}
\sigma(t) = E \biggl \{ \Bigl [ \epsilon(t)-\epsilon_{\rm u}(t)\Bigr ]
-\frac{1}{X}\int_{0}^{\infty} d\omega 
 \int_{0}^{t} \frac{\partial n_{\rm a}}{\partial \tau}(t,\tau,\omega)
\Bigl [\epsilon(\tau)-\epsilon_{\rm u}(\tau) \Bigr ]d\tau \biggr \},
\end{equation}
where $E=\rho\mu X$ is an analog of the Young modulus.
It follows from Eqs. (18), (23), (24) and (27) that
\begin{equation}
A_{2}(t)=\frac{2}{\rho\mu} B\Bigl (\sigma(t), \epsilon(t),\frac{d\epsilon}{dt}(t)\Bigr )
\sigma^{2}(t)\Bigl | \frac{d\epsilon}{dt}(t)\Bigr |.
\end{equation}
According to Eqs. (25) and (28), the functions $A_{1}(t)$ and $A_{2}(t)$
are non-negative for an arbitrary program of loading,
which implies that the Clausius--Duhem inequality (26) is satisfied.

Substitution of Eqs. (3), (14) and (15) into Eq. (27) results in the constitutive 
equation
\begin{eqnarray}
\sigma(t) &=& E \biggl \{ \Bigl [ \epsilon(t)-\epsilon_{\rm u}(t)\Bigr ]
-\kappa_{\rm a}(\epsilon(t)) \int_{0}^{\infty} p(\omega) d\omega 
\nonumber\\
&&\times \int_{0}^{t} \Gamma(\epsilon(t),\omega) 
\exp \biggl [-\int_{\tau}^{t} \Gamma(\epsilon(s),\omega) ds\biggr ]
\Bigl [\epsilon(\tau)-\epsilon_{\rm u}(\tau) \Bigr ]d\tau \biggr \}.
\end{eqnarray}
Given functions $p(\omega)$, $\Gamma_{\rm a}(\epsilon)$ and $\kappa_{\rm a}(\epsilon)$,
the time-dependent response of a semicrystalline polymer 
at isothermal uniaxial loading with small strains is
determined by Eqs. (2), (18) and (29).
For a standard relaxation test with the longitudinal strain $\epsilon^{0}$,
\[
\epsilon(t)=\left \{\begin{array}{ll}
0,  & t<0,\\
\epsilon^{0}, & t\geq 0,
\end{array} \right .
\]
these equations imply that
\begin{equation}
\sigma(t,\epsilon^{0})=C_{1}(\epsilon^{0})-C_{2}(\epsilon^{0})
\int_{0}^{\infty} p(\omega) \biggl [ 1-\exp\Bigl (-\Gamma_{\rm a}(\epsilon^{0})
\exp(-\omega)t\Bigr )\biggr ] d\omega,
\end{equation}
where $\epsilon_{\rm u}^{0}$ is the strain induced by sliding of junctions
and
\begin{equation}
C_{1}(\epsilon^{0})=E(\epsilon^{0}-\epsilon_{\rm u}^{0}),
\qquad
C_{2}(\epsilon^{0})=E(\epsilon^{0}-\epsilon_{\rm u}^{0})
\kappa_{\rm a}(\epsilon^{0}).
\end{equation}
To fit experimental data, we adopt the random energy model \cite{Dyr95} with
\begin{equation}
p(\omega)=p_{0}\exp \biggl [-\frac{(\omega-\Omega)^{2}}{2\Sigma^{2}}\biggr ],
\quad \omega\geq 0,
\qquad
p(\omega)=0,
\quad \omega <0,
\end{equation}
where $\Omega$ and $\Sigma$ are adjustable parameters,
and the pre-factor $p_{0}$ is determined by the condition
\begin{equation}
\int_{0}^{\infty} p(\omega)d\omega =1.
\end{equation}
Given a strain $\epsilon^{0}$, Eqs. (30) and (32) are determined by
5 material constants:
\begin{enumerate}
\item 
an analog of the average potential energy for
rearrangement of strands $\Omega$,

\item
an analog of the standard deviation for distribution of potential energies
$\Sigma$,

\item
the attempt rate for separation of strands from temporary junctions
in active MRs $\Gamma_{\rm a}$,

\item
the coefficients $C_{1}$ and $C_{2}$.
\end{enumerate}
Our aim is to determine these parameters by fitting experimental data
depicted in Figures 3 to 11.

\section{Fitting of observations}

We begin with matching relaxation curves for specimens not subjected to
thermal treratment.
First, we approximate experimental data measured at the strain 
$\epsilon_{2}=0.04$.
This strain is chosen because it is located substantially below
the yield point, on the one hand,
and the testing machine ensures high accuracy of control of the
strain level in the relaxation mode, on the other.

Because the rate of rearrangement, $\Gamma_{\rm a}$, 
and the average potential energy, $\Omega$, are mutually 
dependent [according to Eqs. (30) and (32), the growth 
of $\Omega$ results in an increase in $\Gamma_{\rm a}$], 
we set $\Gamma_{\rm a}=1$ s and approximate 
the relaxation curve by using 4 experimental constants: 
$\Omega$, $\Sigma$, $C_{1}$ and $C_{2}$.
To find these quantities, we fix the intervals 
$[0,\Omega_{\max}]$ and $[0,\Sigma_{\max}]$, 
where the ``best-fit" parameters $\Omega$ and $\Sigma$ 
are assumed to be located, 
and divide these intervals into $J$ subintervals by
the points $\Omega_{i}=i\Delta_{\Omega}$ 
and $\Sigma_{j}=j\Delta_{\Sigma}$  ($i,j=1,\ldots,J$)
with $\Delta_{\Omega}=\Omega_{\max}/J$,
$\Delta_{\Sigma}=\Sigma_{\max}/J$.
For any pair, $\{\Omega_{i},\Sigma_{j} \}$, we evaluate 
the integral in Eq. (30) numerically (by Simpson's method 
with 200 points and the step $\Delta_{\omega}=0.1$).
The pre-factor $p_{0}$ is determined by Eq. (33).
The coefficients $C_{1}=C_{1}(i,j)$ and $C_{2}=C_{2}(i,j)$ 
are found by the least-squares method from the condition
of minimum of the function
\begin{equation}
{\cal J}(i,j)=\sum_{t_{m}} \Bigl [ \sigma_{\rm exp}(t_{m})
-\sigma_{\rm num}(t_{m}) \Bigr ]^{2},
\end{equation}
where the sum is calculated over all experimental points $t_{m}$.
The stress $\sigma_{\rm exp}$ in Eq. (34) is measured 
in the relaxation test, whereas the stress $\sigma_{\rm num}$ 
is given by Eq. (30).
The ``best-fit" parameters $\Omega$ and $\Sigma$ minimize 
the function ${\cal J}$ on the set
$\Bigl \{ \Omega_{i},\; \Sigma_{j} \quad (i,j=1,\ldots, J) \Bigr \}$.
After determining the ``best-fit" values, $\Omega_{i}$ and $\Sigma_{j}$, 
we repeat this procedure for the new intervals $[\Omega_{i-1},\Omega_{i+1}]$
and $[\Sigma_{j-1},\Sigma_{j+1}]$ to ensure good accuracy of fitting.
Figure 4 demonstrates fair agreement between the experimental data 
and the results of numerical simulation 
with $\Omega=4.29$ and $\Sigma=4.34$.

To approximate relaxation curves at other strains, $\epsilon_{k}$,
we fix the constants $\Omega$ and $\Sigma$ found by 
matching observations at $\epsilon_{2}$ 
and fit every relaxation curve by using 3 adjustable parameters: 
$\Gamma_{\rm a}$, $C_{1}$ and $C_{2}$.
These quantities are determined by using a procedure
similar to that employed in the approximation of the relaxation curve 
at $\epsilon_{2}$.
We fix the interval $[0,\Gamma_{\max}]$, 
where the ``best-fit" attempt rate $\Gamma_{\rm a}$ is supposed 
to be located, and divide this interval into $J$ subintervals by the points 
$\Gamma_{i}=i\Delta_{\Gamma}$  ($i=1,\ldots,J$)
with $\Delta_{\Gamma}=\Gamma_{\max}/J$.
For any $\Gamma_{i}$, we calculate the integral in
Eq. (30) numerically and calculate the coefficients $C_{1}=C_{1}(i)$ 
and $C_{2}=C_{2}(i)$ by the least-squares method
from the condition of minimum for function (34).
The ``best-fit" attempt rate minimizes the function ${\cal J}$ 
on the set
$\Bigl \{ \Gamma_{i} \quad (i=1,\ldots, J) \Bigr \}$.
When this ``best-fit"  value, $\Gamma_{i}$, is found,
the procedure is repeated for the new interval 
$[\Gamma_{i-1},\Gamma_{i+1}]$
to ensure an acceptable accuracy of fitting.
Figures 3 to 11 show good agreement between the observations
and the results of numerical analysis.

The above algorithm of fitting is repeated to approximate
the relaxation curves for specimens annealed for 4 and 24 hours.
The ``best-fit" parameters $\Omega$ and $\Sigma$ read
5.70 and 4.88 for samples annealed for 4 h
and 5.19 and 3.80 for specimens annealed for 24 h,
respectively.

For a quasi-Gaussian distribution function (32), 
the parameters $\Omega$ and $\Sigma$ do not coincide
with the average potential energy for detachment of
active strainds, $\Omega_{0}$, and the standard deviation of
potential energies for separation of strands from the network,
$\Sigma_{0}$. 
The latter quantities read
\begin{equation}
\Omega_{0}=\int_{0}^{\infty} \omega p(\omega) d\omega,
\qquad
\Sigma_{0}=\biggl [ \int_{0}^{\infty} (\omega-\Omega_{0})^{2} d\omega 
\biggr ]^{\frac{1}{2}}.
\end{equation}
The dimensionless parameters $\Omega_{0}$, $\Sigma_{0}$ 
and $\xi=\Sigma_{0}/\Omega_{0}$
given by Eq. (35) are listed in Table 1 which
shows that the width of the quasi-Gaussian distribution
(characterized by the ratio $\xi$)
monotonically decreases with annealing time
(however, changes in $\xi$ are rather weak).

For any longitudinal strain $\epsilon_{k}$, the attempt rate, 
$\Gamma_{\rm a}(\epsilon_{k})$, is determined by matching 
an appropriate relaxation curve.
The fraction of active MRs, $\kappa_{\rm a}(\epsilon_{k})$, is found from
Eq. (31),
\[ 
\kappa_{\rm a}(\epsilon_{k})=\frac{C_{2}(\epsilon_{k})}{C_{1}(\epsilon_{k})}.
\]
These quantities are plotted versus strain $\epsilon$ in Figures 12 and 13.
The experimental data are approximated by the phenomenological equations
\begin{equation}
\log \Gamma_{\rm a}=\gamma_{0}+\gamma_{1}\epsilon,
\qquad
\kappa_{\rm a}=k_{0}+k_{1}\epsilon,
\end{equation}
where the coefficients $\gamma_{i}$ and $k_{i}$ are found by the
least-squares method.
Figures 12 and 13 reveal that the viscoelastic behavior 
of specimens not subjected to thermal treatment is rheologically simple
in the sense that the quantities $\Gamma_{\rm a}$ and $\kappa_{\rm a}$
in Eq. (30) are independent of strain.
On the contrary, annealed samples demonstrate the time-dependent 
behavior that is strongly affected by loading: with an increase in
strain, the attempt rate, $\Gamma_{\rm a}$, exponentially decreases 
and the fraction of active MRs, $\kappa_{\rm a}$, linearly grows.

According to Figure 12, the attempt rate, $\Gamma_{\rm a}$, is
strain-independent for non-annealed specimens (however, the scatter 
of the experimental data is rather large). 
A detailed analysis of relaxation curves for non-annealed samples 
\cite{DC02} shows that the attempt rate increases in
the range of strains from $\epsilon=0.005$ to $\epsilon=0.02$
and remains constant at $\epsilon\geq 0.02$.
This implies that the free-volume concept \cite{KE87,LK92,OK95}
is valid for isotactic polypropylene, but the area of its applicability
is confined to relatively small deformations far below the yield strain.

\section{Discussion}

Several approaches were recently proposed to the description of
the viscoplastic behavior of isotactic polypropylene in the
vicinity of the yield point.
Aboulfaraj et al. \cite{AGU95} presumed that plastic slip mechanisms 
had noticeably different features in $\alpha$ and $\beta$ structures.
Karger-Kocsis and Varga \cite{KV96} and Karger-Kocsis et al. \cite{KVE97}
explained toughening of iPP by mechanically-induced $\beta\to\alpha$
transformation of crystallites.
Raab et al. \cite{RKB98} associated the difference in the response
of $\alpha$ and $\beta$ spherulites with different types of
chain folding in lamellae.

Three substantial shortcoming of these concepts should be mentioned:
\begin{enumerate}
\item
they are based on some hypotheses about the difference
in the mechanical behavior of $\alpha$ and $\beta$ crystallites
which have not yet been confirmed experimentally,

\item
these models imply that changes in the stress--strain diagrams
of iPP at annealing are associated with an increase in the
content of $\beta$-polymorph, which contradicts to WAXS 
(wide angle X-ray scattering) diagrams
obtained by Iijima and Strobl \cite{IS00},

\item
these approaches do not expound a pronounsed decrease in the
relaxation rate with strain for annealed samples exhibited in
Figure 12.
\end{enumerate}
The results presented in Figures 1, 12 and 13 may
be explained in terms of the growth of heterogeneity in the
distribution of lamellar strength notwithstanding whether
the fraction of $\beta$ spherulites increases at annealing.

An increase in the low-temperature shoulder of the DSC traces
depicted in Figure 1 (at a practically constant enthalpy of melting)
means that the content of ``weak" crystalline lamellae (that melt
at relatively low temperatures) noticeably grows.
These ``weak" lamellae may be associated with subsidiary
lamellae in $\alpha$ spherulites.

Using a thermo-mechanical analogy, one can speculate that
the growth of the fraction of thermally weak lamellae at annealing
is tantamount to an increase in the concentration
of lamellae which can be easily fragmented under stretching.
The latter reflects an increase in structural disorder of
$\alpha$ crystallites at annealing far below the melting point
observed as substantial variations in the relative intensity
of Bragg reflections \cite{ARR00}.

On the other hand, annealing of isotactic polypropylene induces
thickening of radial lamellae, which results in the growth of 
elastic moduli of the polymer.
This implies that at relatively small strains ($\epsilon<0.02$ to 0.04,
when fragmentation of lamellae does not occur), annealing of iPP 
leads to an increase in the longitudinal stress (which is 
demonstrated in Figure 2).
At higher strains, the coarse slip starts in weak lamellae 
of specimens annealed for 24 h (which are less homogeneous
than those annealed for 4 h), whereas no lamellar fragmentation
takes place in samples annealed for 4 h.
As a consequence, the stress--strain curve 2 is located higher
that the curve 3 in Figure 2.
With an increase in strain, lamellar fragmentation occurs
in both specimens, which implies that the stress--strain
curves for annealed specimens practically coincide at 
strains exceeding the yield point $\epsilon_{\rm y}=0.13$.

Further increase in strain in the post-yield region results 
in total fragmentation of lamellae that can be broken 
at a given stress intensity in specimens not subjected 
to thermal treatment.
Because homogeneous (along a specimen) crystal slip 
becomes impossible, spatial heterogeneity in the distribution
of stresses arises at the macro-level, which leads to
necking of the specimens at $\epsilon_{\rm n}=0.18$.
For the samples annealed for 4 and 24 hours, 
the concentration of ``weak" lamellae is higher, 
which implies that the strains corresponding
to the total fragmentation of weak lamellae 
and transition to the spatially heterogeneous deformation 
of specimens exceed that for the non-annealed material 
($\epsilon_{\rm n}=0.25$ and $\epsilon_{\rm n}>0.30$,
respectively).

Noticeable fragmentation of lamellae in spherulites implies that
some amorphous regions  are released (whose deformation was 
previously screened by surrounding lamellae in non-broken crystallites),
which results in an increase in the fraction of active MRs, 
$\kappa_{\rm a}$, with strain.
This release of restricted amorphous phase is substantially less 
pronounced in the specimens not subjected to thermal 
treatment (curve 1 in Figure 13) compared with the annealed specimens 
(curves 2 and 3 in Figure 13).
Three features of the curves depicted in Figure 13 are worth to
be mentioned:
\begin{enumerate}
\item
At small strains (less than 0.04), the concentration of active
MRs monotonically decreases with annealing time.
This phenomenon may be attributed to an increase in the fraction
of amorphous regions whose mobility is restricted by
surrounding lamellae at annealing (driven by development of
subsidiary lamellae).

\item
The content of active meso-domains linearly grows with
strain in annealed specimens with the rate that
is practically independent of the annealing time.
This may be explained by changes in the micro-structure
of spherulites at annealing: although the rate of lamellar 
fragmentation is assumed to be higher in iPP annealed
for 24 h, the amount of amorphous phase released at any
fragmentation act is smaller than in iPP annealed for 4 h.

\item
Curves 2 and 3 intersect curve 1 in the region between
$\epsilon=0.08$ and $\epsilon=0.12$, i.e., in the close
vicinity of the yield point for non-annealed specimens
(see Figure 2).
\end{enumerate}
With reference to Coulon et al. \cite{CCG98} and 
Raab et al. \cite{RKB98}, we suppose that under stretching
lamellae are fragmented into small aligned blocks 
that serve as extra physical cross-links in amorphous meso-domains.
According to the concept of transient networks \cite{TE92},
an increase in the concentration of permanent
cross-links results in a decrease in the net rate of rearrangement.
This rate is characterized in the model by the attempt
rate, $\Gamma_{\rm a}$, which is considered as an
average (over MRs) rate of detachment of active strands 
from their junctions.
This conclusion is fairly well confirmed by the results
depicted in Figure 12: 
the attempt rate is practically independent of strain for
the specimens not subjected to thermal pre-treatment, and
$\Gamma_{\rm a}$ exponentially decreases with
strain for the annealed samples.
Dispite apparent similarity in the slopes of curves 2 and 3
in Figure 12, it is rather difficult to assert that the kinetics
of the strain-induced decrease in the attempt rate is independent
of annealing time because of the large scatter of
data for the specimens annealed for 24 h.
It is worth noting that the attempt rates were determined in
Section 5 from the condition $\Gamma_{\rm a}=1$ s
at $\epsilon=0.04$, which implies that their values 
cannot be directly compared for specimens annealed for
different amounts of time (because these samples
have different distributions of potential energies
for separation of active strands from temporary nodes, 
see Table 1).
 
\section{Concluding remarks}

Constitutive equations have been derived for the time-dependent behavior
of semicrystalline polymers at isothermal loading with small strains.
A mean-field approach is employed to develop stress--strain relations: 
a complicated micro-structure of isotactic polypropylene
is replaced by an equivalent transient network of macromolecules bridged by
junctions (physical cross-links, entanglements and crystalline lamellae).
The network is assumed to be strongly inhomogeneous, and it is
thought of as an ensemble of meso-regions with various potential energies 
for separation of strands from temporary nodes.

The viscoelastic response of a semicrystalline polymer is ascribed 
to separation of active strands from temporary junctions
and merging of dangling strands to the network in active 
meso-domains.
Rearrangement of strands is modelled as a thermo-mechanically 
activated process whose rate is given by the Eyring equation
with a strain-dependent attempt rate.

The viscoplastic response is described by slippage of junctions 
with respect to their positions in the bulk material.
The rate of sliding is assumed to be proportional to 
the macro-stress in a specimen.

Three series of tensile relaxation tests have been performed 
on isotactic polypropylene at ambient temperature.
In the first series, injection-molded samples are used without
thermal pre-treatment.
In the second series, the samples are annealed at  130 $^{\circ}$C
for 4 h, and in the last series, the speciments are annealed
for 24 h at the same temperature.
Adjustable parameters in the stress--strain relations 
are found by fitting observations in the range of strains
from 0.02 to 0.18.
The following conclusions are drawn from the analysis of experimental data:
\begin{enumerate}
\item
The relaxation spectrum of iPP (which is determined by the distribution
function, $p(\omega)$, for potential energies of detachment of active 
strands from their junctions)
is not affected by mechanical factors, but is altered at annealing.

\item
The attempt rate, $\Gamma_{\rm a}$, for separation of active strands 
from temporary nodes is practically independent of strain 
for specimens not subjected to thermal treatment, 
and it exponentially decreases with strain for annealed samples.

\item
The relaxation strength (which is characterized by the content of
active meso-regions $\kappa_{\rm a}$) is independent of strain
for non-annealed specimens, and it linearly increases with strain for
annealed samples.
\end{enumerate}
These findings are qualitatively explained based on the hypothesis
that the distribution of strengths of crystalline lamellae in iPP  
is noticeably broadened at annealing, which results not only in
thickening of lamellae, but also in the growth of ``weak" (subsidiary)
lamellae that are easily fragmented either by heating
or by mechanical loading.
DSC measurements provide some confirmation for this assumption.

\newpage
\section*{List of figures}

{\bf Figure 1:} DSC melting curves for a non-annealed 
specimen (unfilled circles),
a specimen annealed for 4 h (filled circles) 
and a specimen annealed for 24 h (triangles).
Symbols: experimental data
\vspace*{2 mm}

\noindent
{\bf Figure 2:}
The longitudinal stress $\sigma$ MPa
versus strain $\epsilon$ in tensile tests with 
the cross-head speed 5 mm/min.
Symbols: experimental data.
Unfilled circles: a virgin specimen;
filled circles: a specimen annealed for 4 h;
triangles: a specimen annealed for 24 h;
asterisks: stresses at the beginning of relaxation tests 
at various strains
\vspace*{2 mm}

\noindent
{\bf Figure 3:}
The longitudinal stress $\sigma$ MPa
versus time $t$ s in a tensile relaxation test 
at $\epsilon=0.02$.
Symbols: experimental data.
Solid lines: results of numerical simulation.
Curve 1: a virgin specimen;
curve 2: a specimen annealed for 4 h;
curve 3: a specimen annealed for 24 h
\vspace*{2 mm}

\noindent
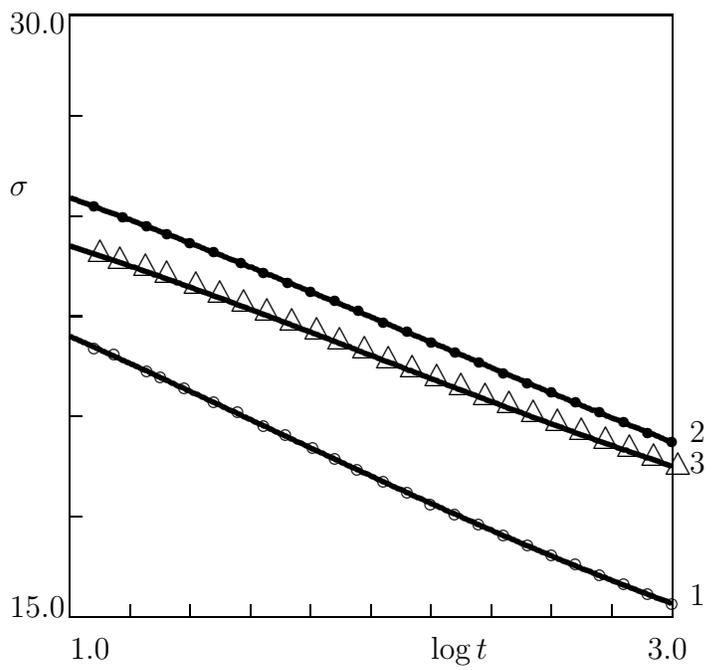
{\bf Figure 4:}
The longitudinal stress $\sigma$ MPa
versus time $t$ s in a tensile relaxation test 
at $\epsilon=0.04$.
Symbols: experimental data.
Solid lines: results of numerical simulation.
Curve 1: a virgin specimen;
curve 2: a specimen annealed for 4 h;
curve 3: a specimen annealed for 24 h
\vspace*{2 mm}

\noindent
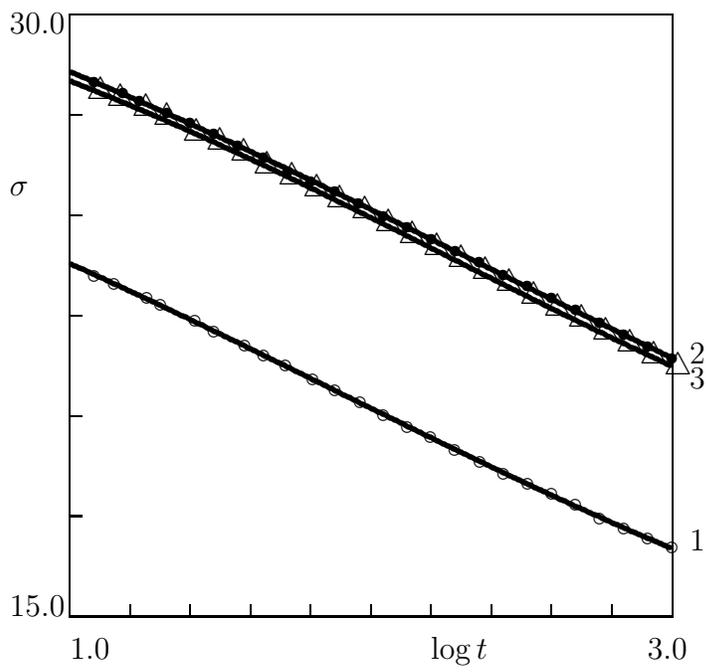
{\bf Figure 5:}
The longitudinal stress $\sigma$ MPa
versus time $t$ s in a tensile relaxation test 
at $\epsilon=0.06$.
Symbols: experimental data.
Solid lines: results of numerical simulation.
Curve 1: a virgin specimen;
curve 2: a specimen annealed for 4 h;
curve 3: a specimen annealed for 24 h
\vspace*{2 mm}

\noindent
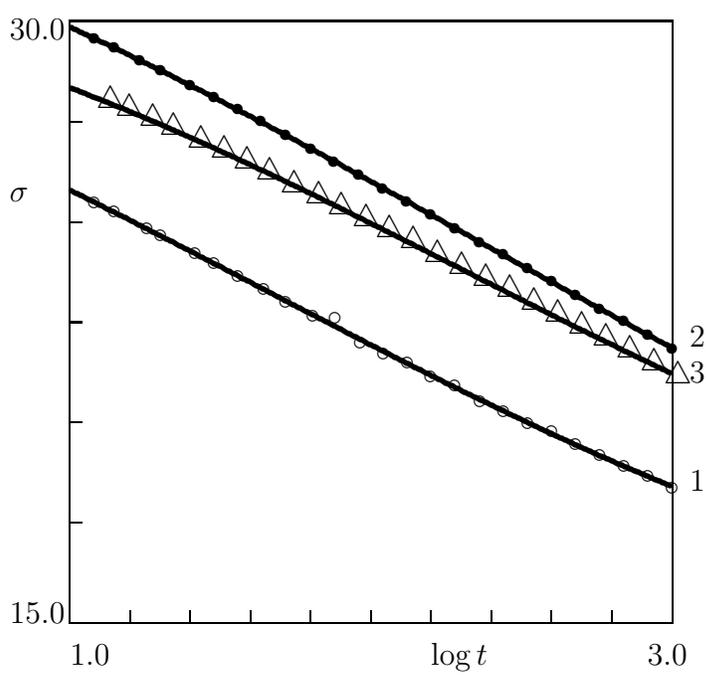
{\bf Figure 6:}
The longitudinal stress $\sigma$ MPa
versus time $t$ s in a tensile relaxation test 
at $\epsilon=0.08$.
Symbols: experimental data.
Solid lines: results of numerical simulation.
Curve 1: a virgin specimen;
curve 2: a specimen annealed for 4 h;
curve 3: a specimen annealed for 24 h
\vspace*{2 mm}

\noindent
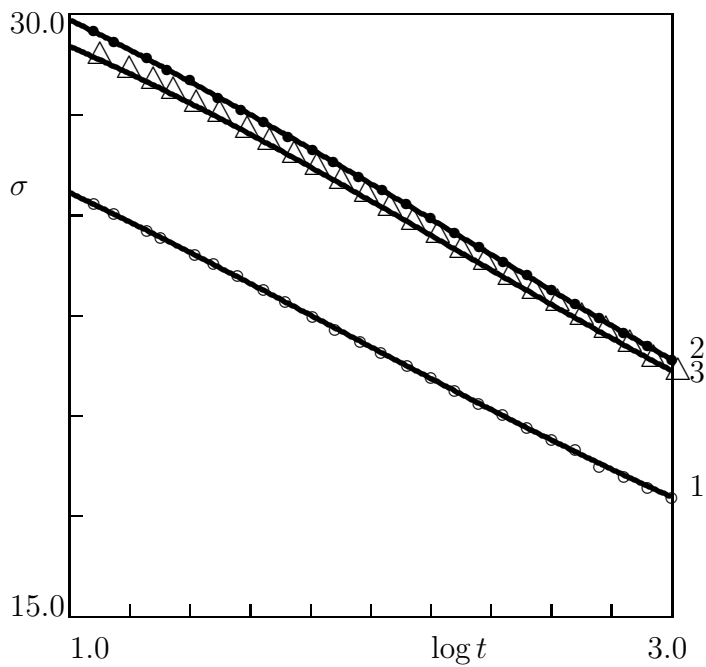
{\bf Figure 7:}
The longitudinal stress $\sigma$ MPa
versus time $t$ s in a tensile relaxation test 
at $\epsilon=0.10$.
Symbols: experimental data.
Solid lines: results of numerical simulation.
Curve 1: a virgin specimen;
curve 2: a specimen annealed for 4 h;
curve 3: a specimen annealed for 24 
\vspace*{2 mm}

\noindent
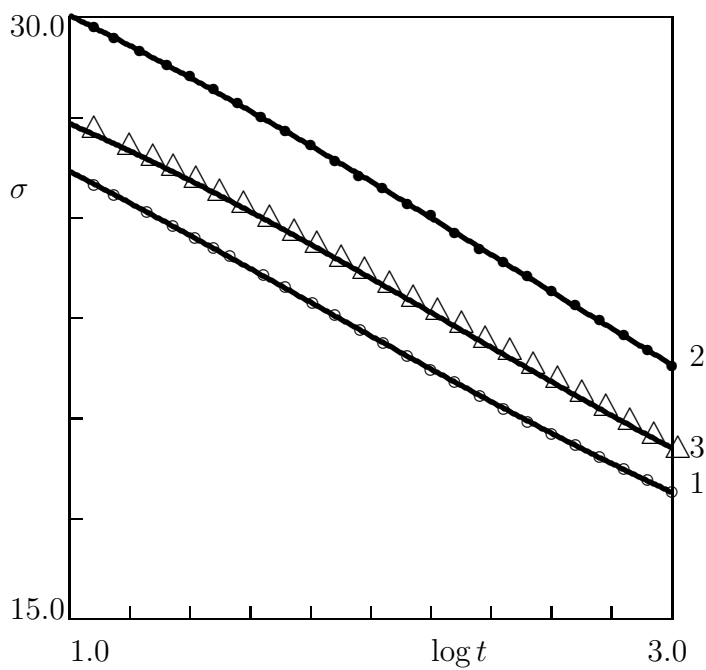
{\bf Figure 8:}
The longitudinal stress $\sigma$ MPa
versus time $t$ s in a tensile relaxation test 
at $\epsilon=0.12$.
Symbols: experimental data.
Solid lines: results of numerical simulation.
Curve 1: a virgin specimen;
curve 2: a specimen annealed for 4 h;
curve 3: a specimen annealed for 24 h
\vspace*{2 mm}

\noindent
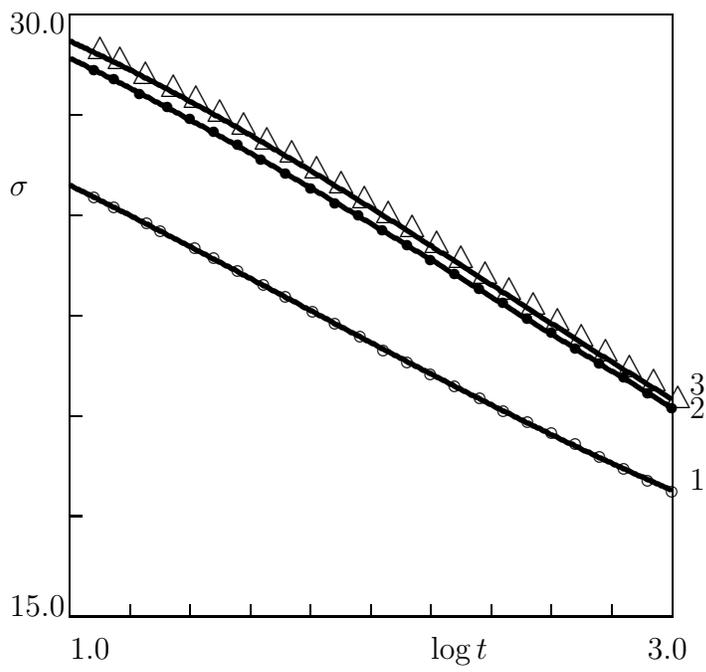
{\bf Figure 9:}
The longitudinal stress $\sigma$ MPa
versus time $t$ s in a tensile relaxation test 
at $\epsilon=0.14$.
Symbols: experimental data.
Solid lines: results of numerical simulation.
Curve 1: a virgin specimen;
curve 2: a specimen annealed for 4 h;
curve 3: a specimen annealed for 24 h
\vspace*{2 mm}

\noindent
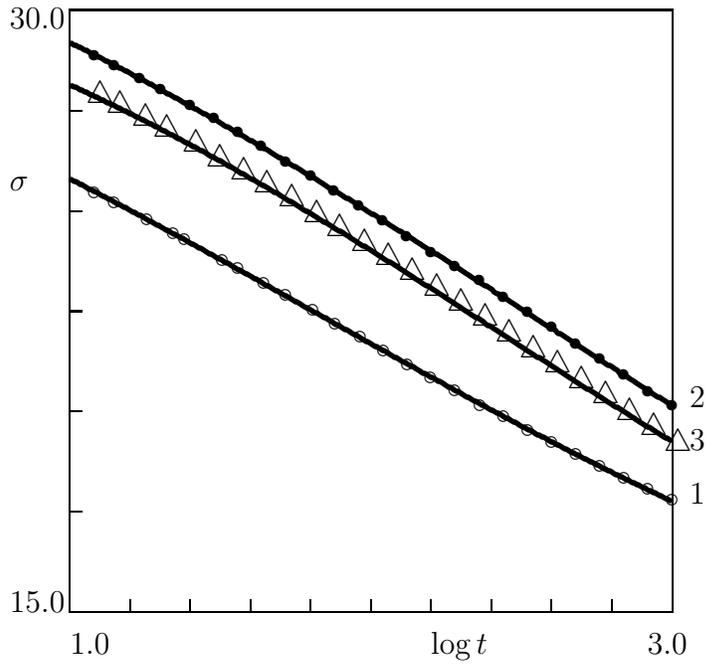
{\bf Figure 10:}
The longitudinal stress $\sigma$ MPa
versus time $t$ s in a tensile relaxation test 
at $\epsilon=0.16$.
Symbols: experimental data.
Solid lines: results of numerical simulation.
Curve 1: a virgin specimen;
curve 2: a specimen annealed for 4 h;
curve 3: a specimen annealed for 24 h
\vspace*{2 mm}

\noindent
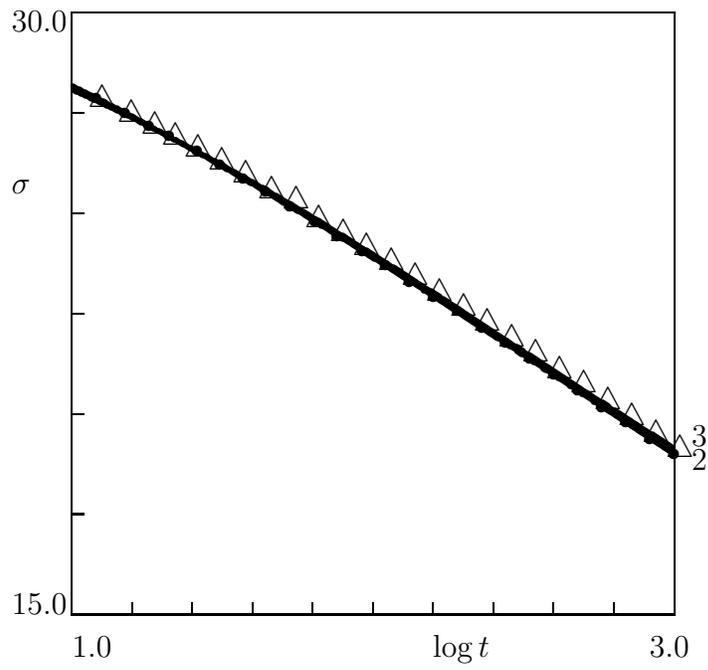
{\bf Figure 11:}
The longitudinal stress $\sigma$ MPa
versus time $t$ s in a tensile relaxation test 
at $\epsilon=0.18$.
Symbols: experimental data.
Solid lines: results of numerical simulation.
Curve 2: a specimen annealed for 4 h;
curve 3: a specimen annealed for 24 h
\vspace*{2 mm}

\noindent
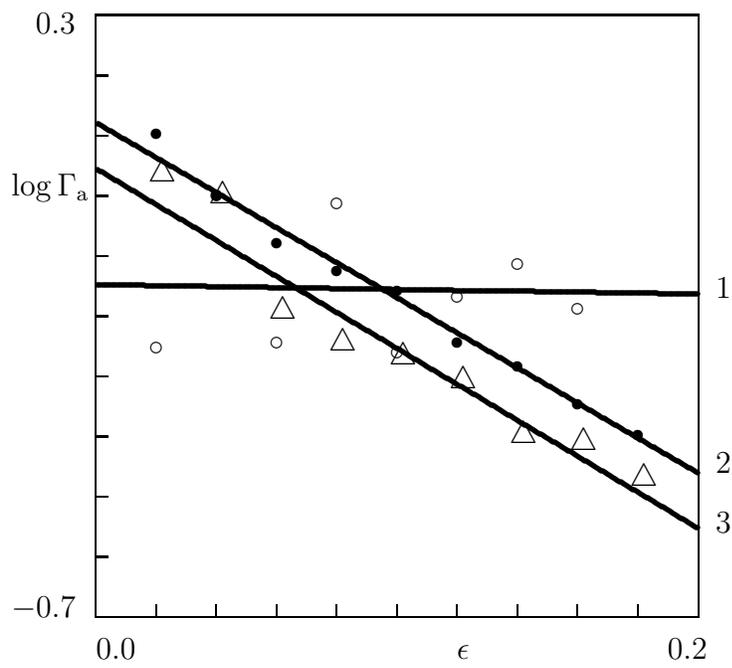
{\bf Figure 12:}
The attempt rate $\Gamma_{\rm a}$ s$^{-1}$
versus strain $\epsilon$ in tensile relaxation tests.
Symbols: treatment of observations.
Unfilled circles: virgin specimens;
filled circles: specimens annealed for 4 h;
triangles: specimens annealed for 24 h.
Solid lines: approximation of the experimental data by Eq. (36).
Curve 1: $\gamma_{0}=-0.15$, $\gamma_{1}=-0.08$;
curve 2: $\gamma_{0}=0.12$, $\gamma_{1}=-2.91$;
curve 3: $\gamma_{0}=0.05$, $\gamma_{1}=-2.99$
\vspace*{2 mm}

\noindent
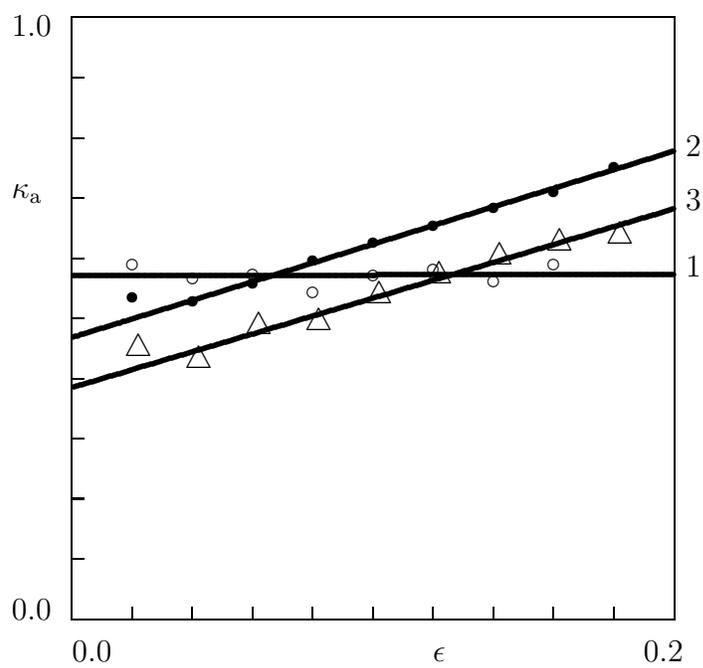
{\bf Figure 13:}
The concentration of active MRs $\kappa_{\rm a}$
versus strain $\epsilon$ in tensile relaxation tests.
Symbols: treatment of observations.
Unfilled circles: virgin specimens;
filled circles: specimens annealed for 4 h;
triangles: specimens annealed for 24 h.
Solid lines: approximation of the experimental data by Eq. (36).
Curve 1: $k_{0}=0.57$, $k_{1}=0.01$;
curve 2: $k_{0}=0.47$, $k_{1}=1.55$;
curve 3: $k_{0}=0.39$, $k_{1}=1.49$
\newpage

\begin{table}[t]
\begin{center}
\caption{Adjustable parameters $\Omega_{0}$, $\Sigma_{0}$ and $\xi$
at various annealing times $t_{\rm a}$ h}
\vspace*{6 mm}


\end{center}
\vspace*{5 mm}

\caption{}
\end{figure}

\begin{figure}[tbh]
\begin{center}
%
\end{center}
\vspace*{5 mm}

\caption{}
\end{figure}

\begin{figure}[tbh]
\begin{center}
%
\end{center}
\vspace*{5 mm}

\caption{}
\end{figure}

\begin{figure}[tbh]
\begin{center}
%
\end{center}
\vspace*{5 mm}

\caption{}
\end{figure}
\end{document}